\documentclass[graybox]{svmult}

\usepackage{mathptmx}       
\usepackage{helvet}         
\usepackage{courier}        
\usepackage{type1cm}        
\usepackage{amsmath}
\usepackage{amssymb}

\usepackage{makeidx}         
\usepackage{graphicx}        
\usepackage{multicol}        
\usepackage[bottom]{footmisc}

\makeindex             

\begin{document}
\title{Direction and Constraint in Phenotypic Evolution: Dimension Reduction and Global Proportionality in Phenotype Fluctuation and Responses}
\author{Kunihiko Kaneko 1) and Chikara Furusawa 2)}
\institute{1) Research Center for Complex Systems Biology, Universal Biology Institute, University of Tokyo, 3-8-1 Komaba, Tokyo 153-8902, Japan 
2) Center for Biosystems Dynamics Research, RIKEN, 6-2-3 Furuedai, Suita, Osaka 565-0874;\\ and Universal Biology Institute, University of Tokyo, 7-3-1 Hongo, Tokyo 113-0033, Japan }

\maketitle
\abstract{
A macroscopic theory for describing cellular states during steady-growth is presented, which is based on the consistency between cellular growth and molecular replication, as well as the robustness of phenotypes against perturbations. 
Adaptive changes in high-dimensional phenotypes were shown to be restricted within a low-dimensional slow manifold, from which a macroscopic law for cellular states was derived, which was confirmed by adaptation experiments on bacteria under stress. Next, the theory was extended to phenotypic evolution, leading to proportionality between phenotypic responses against genetic evolution and environmental adaptation. The link between robustness to noise and mutation, as a result of robustness in developmental dynamics to perturbations, showed proportionality between phenotypic plasticity by genetic changes and by environmental noise. Accordingly, directionality and constraint in phenotypic evolution was quantitatively formulated in terms of phenotypic fluctuation and the response against environmental change. The evolutionary relevance of slow modes in controlling high-dimensional phenotypes is discussed.
}

\section{Introduction}

In a chapter in the previous volume of Evolutionary Systems Biology \cite{ESB}, we discussed the evolutionary fluctuation-response relationship, which states that if phenotypic variance due to noise is high, then evolution rapidly occurs. This suggests a correlation between short-term phenotypic dynamics and long-term evolutionary responses. This, in some sense, is a quantitative expression of Waddington's genetic assimilation \cite{Waddington}.
 
Can we push this viewpoint forward to determine the direction of phenotypic evolution in a high-dimensional phenotypic space (i.e., with a large degree of freedom components)? Can one predict which traits are likely to evolve among many components before evolution progresses?
 
To answer the question, we first investigate the characteristics of responses of phenotypes with a large degree of freedom. We review the general relation in phenotypic responses of cells over many components, demonstrating that global proportionality exists among all logarithmic changes in concentrations against adaptation to different environmental conditions. We first discuss this proportionality as a general consequence of steady exponential growth cell, following \cite{Mu}. We show that when a cell grows and divides while maintaining its composition, the abundances of each component increases at the same rate; this constraint supports the global proportional relationship. 
 
However, the growth-rate constraint is not enough to explain the experimental observations. Global proportional changes across all components are confirmed even across many different environmental conditions. This result cannot be explained by the constraint of steady-growth alone.
We will see that another important constraint, imposed by the robustness in a phenotypic state shaped throughout evolution, is essential. From this evolutary robustness, phenotypes can change mostly along only one or in a few dimensions, although the original phenotypic space is high-dimensional as a consequence of the huge diversity in the components of cells. Based on \cite{CFKK-Dominant}, We demonstrate this evolutionary dimension-reduction both through theory and simulations, whereas its consequence is consistent with experimental observations.
 
This constraint in phenotypic changes is extended to changes that occur in evolution. We demonstrate that long-term phenotypic changes via evolution and short-term changes via adaptation are highly correlated. Global proportionality in the phenotypic changes by environmentally induced adaptation and those by genetically induced evolution is confirmed across all components, both in simulations and laboratory evolution experiments \cite{CFKK-Interface}. 
 
In contrast, response and fluctuation are {\sl two sides of the same coin},
as has been demonstrated by statistical mechanics (see also the first volume of Evolutionary Systems Biology \cite{ESB}). Hence, a similar correlation in concentration fluctuations is expected across all components. Indeed, we demonstrate a proportional relationship between fluctuations by gene mutation and those by noise over the concentration of all components. 
Recall that the variances in each trait (phenotype) due to genetic variation are proportional to the evolution rate of the trait according to the fundamental theorem of natural selection by Fisher \cite{Fisher}. Hence, the evolution rate of each trait is correlated with its variance by noise. This variance is predetermined before mutation and selection. This means that the evolutionary potential of each trait is determined in advance by the phenotype changeability which is affected by environmental variation or noise before genetic changes occur. This enables the prediction of phenotypic evolution. Among the high-dimensional phenotypic space, evolution progresses along the direction in which the variation by the noise or environmental response is larger, which is predetermined before mutation. Although genetic variation itself is random and undirected, phenotype evolution tends to show directionality.

\section{Constraint in a Steady-Growth system: Global Proportionality Law}

To describe changes in the cellular state in response to environmental changes, we introduce a simple theory by assuming that cells undergo steady-growth. When a cell grows and reproduces in this steady state, all components, e.g., expressed proteins, must be approximately doubled \cite{book,Zipf}. 
 
Consider a cell consisting of $M$ chemical components. In the cellular state under steady-growth conditions, the cell number increases exponentially over time, as does the cell volume $V$, as given by $dV/dt=\mu V$. In a steady-growth cell, the abundance of all components increases at the same rate, preserving the concentration of each component during the cell cycle.
 
To formulate the constraint for steady-growth, let
 us denote the concentration $x_i(>0)$ for each component $i=1,\cdots,M$. The cellular state is represented as a point in an $M$-dimensional state space.
Here, each component $i$ is synthesized or decomposed relative to other components at a rate $f_i(\{ x_j \})$, for instance, by the rate-equation in chemical kinetics. 
Additionally, all concentrations are diluted by the rate $(1/V)(dV/dt)=\mu$, so that the time-change of a concentration is given by
\begin{equation}
dx_i/dt=f_i(\{x_j\}) -\mu x_i.
\end{equation}
For convenience, let us denote $X_i=\log x_i$, and $f_i=x_i F_i$. Then, Eq. (1) can be written as $dX_i/dt=F_i(\{X_j\}) -\mu$,
which assumes that $x_i\neq 0$, i.e., all components exist. 
Then, the stationary state is given by the fixed-point solution $F_i(\{X^*_j \}) =\mu$ for all $i$.
 
In response to environmental changes, the term $F_i(\{X_j\})$ and growth rate $\mu$ change, 
as does each concentration $x_i^*$; however, the $M-1$ conditions $F_1=F_2=\cdots =F_M$ must be satisfied.
Thus, a cell must follow a 1-dimensional curve in the $M$-dimensional space (see Fig. 1) under a given change in the environmental conditions (e.g., against changes in stress strength).
Now, consider intracellular changes in response to environmental changes.
Here each environmental change given by a type of stress $a$ is parametrized by a single continuous parameter $E^a$ (such as the temperature, degree of nutrient limitation, etc.).
Using this parameterization $E^a$, the steady-growth condition leads to
\begin{math}
F_i(\{ X^*_j(E^a) \},E^a)=\mu(E^a).
\end{math}

\noindent
\begin{figure}[ht]
\begin{center}
\includegraphics[width=4.5cm]{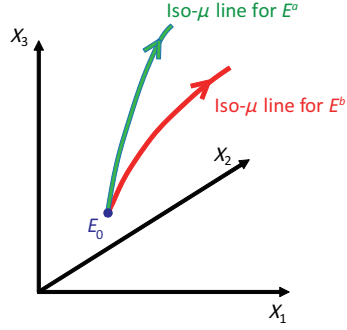}
\end{center}
\caption{
Schematic representation of our theoretical analysis: Changes in gene expression in a high-dimensional state space under different perturbations $E^a$ and $E^b$ are presented. Upon a given environmental change, the phenotypic changes in component concentrations follow a curve satisfying the constraint showing that the growth rates of all components are identical, i.e., an iso-$\mu$ line $F_1=F_2=\cdots =F_M$, in an $M$-dimensional state space. For a different environmental vector, the locus in the state space follows a different iso-$\mu$ line.
}
\end{figure}

We consider the parameter change from $E_0$ to $E$, where each $X_j^*$ changes from $X_j^*$ at $E_0$, to $X_j^*+\delta X_j$, which is accompanied by a change from $\mu$ to $\mu+\delta \mu$. Assuming a gradual change in the dynamics $x_j$, we introduce a partial derivative of $F_i(\{ X^*_j(E) \})$ by $X_j$ at $E=E_0$, which gives the Jacobi matrix $J_{ij}$. Assuming that the environmental change is small and that phenotypic changes are sufficiently small and follow only the linear term in $\delta X_j$, we obtain
\begin{equation}
\sum_j J_{ij} \delta X_j (E) + \gamma_i \delta E =\delta \mu (E)
\end{equation}
with
\begin{math}
\gamma_i \equiv \frac{\partial F_i}{\partial E}.
\end{math}
Under our linear conditions, $\delta \mu \propto \delta E$, so that $\delta \mu= \alpha \delta E$ holds for a constant $\alpha$. Accordingly, we obtain
$\sum_j J_{ij} \delta X_j (E) =\delta\mu (E)(1-\gamma_i/\alpha)$.
Hence, \begin{math} \frac{\sum_j J_{ij} \delta X_j (E) }{\delta\mu (E)}= \frac{\sum_j J_{ij} \delta X_j (E') }{\delta\mu (E')}
\end{math}
so that 
\begin{equation}
\frac{\delta X_j(E)}{\delta X_j (E')} =\frac{\delta \mu (E)}{\delta \mu (E')}
\end{equation}
\noindent is obtained overall $j$.
The formula can be compared with experimental observations. Note that it can be applied to any component. For example, one can use the concentration of either mRNA or protein, depending on the available experimental data.

\section{Experimental Confirmation}

To explore the relationship between changes in global gene expression
and growth rate, we analyzed transcriptome data of {\sl Escherichia
coli} obtained under three environmental conditions: osmotic
stress, starvation, and heat stress, as presented in \cite{Matsumoto}.
 In the experiments, the cells were initially cultured under minimal medium at 37$^\circ$C.
After the transient response to the introduction of a given stress, the cells
were harvested when the growth rate reached a constant value. The data were taken only from an exponentially growing steady state. For each stress condition, three
levels of stresses (s = high, medium, low) were used, so that the
absolute expression levels, represented by $x_j$ for $j$-th gene, were
measured over $3\times3$ conditions.

Through transcriptome analysis under different environmental conditions, we calculated the change in gene expression levels between the original state and for a system experiencing environmental stress. We investigated the difference in gene expression using a log-scale ($X_j=\log x_j$), that is $\delta X_j(E)=X_j(E)-X_j^O$ (i.e., $\log(x_j(E)/x_j^O)$) for genes $j$, where $E$ represents a given environmental condition and $X_j^O$ represents the log-transformed gene expression level under the original condition \cite{Mu}.
 
To examine the validity of the theory for global changes in expression induced by the environmental stresses, we plotted the relationship between the differences in expression $(\delta X_j(E^a_{s_1}),\delta X_j(E^a_{s_2}))$ in Fig. 2(a)-(c) for $s_1=$ low and $s_2=$ medium, where $a$ is either osmotic, heat, or starvation stress. A common proportionality was observed in concentration changes across most mRNA species, which is consistent with the theory \cite{Mu}.
 
According to our theory, the proportion coefficient in the expression level should agree with the growth rate. Here, for each condition, the change in the growth rate $\delta \mu (E^a_s)$ was also measured ($a$ is either osmotic, heat, or starvation stress). 
The slope fitted from the data agrees well with the common ratio $\delta X_j(E^a_{s_1})/\delta X_j(E^a_{s_2})$.

\begin{figure}[ht]
\begin{center}
\includegraphics[width=\linewidth]{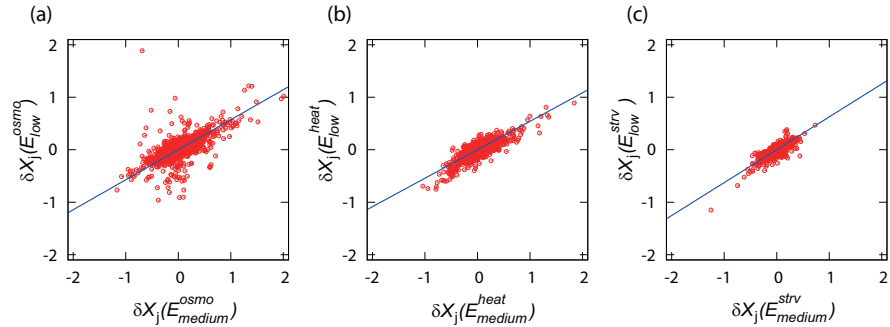}
\end{center}
\caption{Examples of the relationship between changes in gene expression $\delta X_j(E^a_{s_1})$ and $\delta X_j(E^a_{s_2})$ for genes in {\sl E. coli}. $\delta X_j$ represents the difference in the logarithmic expression level of a gene $j$ between non-stressed and stressed conditions, where $s_1$ and $s_2$ represent two different stress strengths, i.e., low and medium. (a), (b), and (c) show the plot for $a=$ osmotic stress, heat stress, and starvation, respectively. 
The fitted line indeed agrees well with that expected from the growth-rate change in Eq.(3).
Reproduced from \cite{Mu}.
}
\end{figure}

In this respect, the theory based on the steady-growth state and linearization of changes in stress works well for analyzing transcriptome changes in bacteria. Indeed, relevance of growth-rate to global trend in transcriptome changes was noted in several experiments \cite{Regenberg,ODuibhr,Botstein,Alon}, and the formulation in the last section can provide a step to understand such global trend. Here, however, the steady-growth theory is not sufficient.
First, the global proportionality is satisfied even under a stress condition that reduces the growth rate to below $20\%$ or as compared to the standard. Such expansion of the linear regime is beyond the simple theory. The other important point missed by this steady-growth theory will be discussed in the next section.

\section{Global Proportional Changes in Gene Expression Beyond the Simple Theory }

Until now, we have compared the responses against a given type of environmental condition with different strengths. In general, possible environmental changes are described by a vector as ${\bf E}$. In the study, the changes are given by ${\bf E}=\lambda^a {\bf e}^{a}$ with different strengths $\lambda^a$ by fixing $ {\bf e}^{a}$.
However, one can also compare expression changes across different types of stress conditions. In this case, the environmental change ${\bf E}$ is no longer represented by a scalar variable, and the responses against the environmental changes given by different vectors ${\bf e}^{a}$ and ${\bf e}^{b}$ must be compared.
 
However, the above theory cannot predict the common proportionality. This is because each one-dimensional curve upon a given type of environmental change is generally located along a different direction in the state space (see Fig. 1). This can also be understood using Eqs. (1)-(3) in Sec. 2. To compare the responses against different types of environmental changes using Eqs. (2) and (3), one needs 
$\gamma_i^a \equiv \frac{\partial F_i}{\partial \lambda^a}$, which depends on the type of environmental (stress) condition $a$. Hence, rather than Eq. (3), we obtain
\begin{equation}
\frac{\delta X_j ({\bf E}^a)}{\delta X_j ({\bf E}^b)} =\frac{\delta \mu ({\bf E}^a)}{\delta \mu ({\bf E}^b)}.
\frac{ \sum_i L_{ji}(1-\gamma_i^a/\alpha^a) } {\sum_i L_{ji} (1-\gamma_i^b/\alpha^b)}
\end{equation}
Then, because $\gamma_i^a \neq \gamma_i^b$, in general, the proportionality cannot be determined as in Eq. (3).
 Although the theory cannot predict the simple proportional relationship, one can plot the experimental data as Fig.2, even across different conditions for osmotic pressure, heat, and starvation. An example of such  plot is given in Fig. 3(a) (see also \cite{Mu}). To further support the proportionality, we plotted $\delta X_j({\bf E})$ for the changes in concentrations of thousands of proteins (rather than mRNA) under different conditions by using proteome analysis \cite{Heinemann} in Fig. 3(b).
Interestingly, in both cases, a strong correlation between $\delta X_j({\bf E}^a)$ and $\delta X_j({\bf E}^b)$ was still observed for all components $j$, even under different environmental conditions. Although more genes deviated from the common proportionality, lowering the correlation coefficients as compared with those for the same type of stress, the global proportionality still held for most genes (Note that Fig. 3(a) and (b) show more than 1000 points, and so that a single line fits most of these points). Thus, the global proportionality is still valid.
Further, as shown in Fig. 3(c), the slope approximately agrees with the rate of growth change, as in Eq. (3). Additionally, other data suggested such a correlation in mRNA abundance under different environmental conditions \cite{Alon,Braun}. 

\begin{figure}[ht]
\begin{center}
\includegraphics[width=11cm]{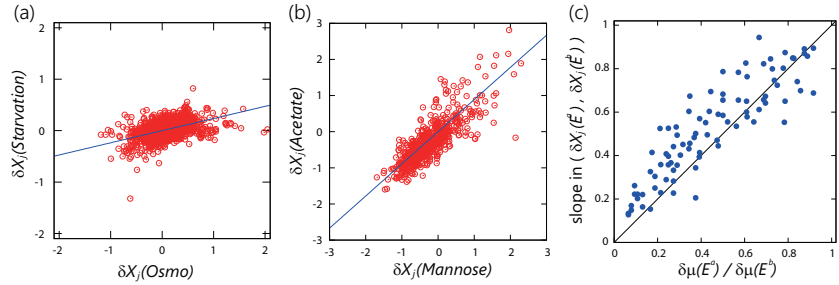}
\end{center}
\caption{
Common proportionality in {\it E. coli} gene expression changes.
(a) shows an example of the relation between mRNA expression changes $\delta X_j({\bf E}^a)$ and $\delta X_j({\bf E}^b)$, where $a$ and $b$ are osmotic stress and starvation, respectively.  
(b) shows an example of the relation between protein expression changes in the mannose carbon source and acetate carbon source conditions. The blue lines in (a) and (b) represent the slope calculated by the ratio of observed growth rate changes. 
(c) Relation between the slope of the change in protein expression and the change in the growth rate. The abscissa represents $\delta\mu({\bf E}^a) / \delta\mu({\bf E}^b)$, whereas the ordinate is the slope in $\delta X_j({\bf E}^a) / \delta X_j({\bf E}^b)$. The slope was obtained by fitting the protein expression data. Reproduced from \cite{CFKK-Dominant}.
}
\end{figure}

Because gene expression dynamics are very high-dimensional, this correlation suggests that a strong constraint exists in adaptive changes in expression dynamics that cannot be explained by the simple theory assuming only steady-growth. The global proportionality is beyond the scope of the simple theory presented in Eqs. (2) and (3). Thus, this proportionality is not generic in any dynamical systems satisfying only steady-growth. 
 
In summary, two questions remain: how to evaluate a broad range of linearity regime and evaluating proportionality under different environmental conditions. To answer these questions, some factor other than steady-growth must be evaluated.

Of course, cells are not only constrained by steady-growth but also are a product of evolution. Through evolution, cells can efficiently and robustly reproduce themselves under external conditions.
Therefore, the above two features may result from evolution. In the next section, we examine the validity of the hypothesis that evolutionary robustness constrains intracellular dynamics to exhibit global proportionality in the adaptive changes of many components.

\section{Emergence of Global Proportionality through Evolution: Formation of a Dominant Mode}

\subsection{Catalytic-reaction network model for numerical evolution}

The above hypothesis regarding the consequence of evolutionary robustness is difficult to evaluate experimentally, as the experimentally available data are only from organisms that currently exist as a result of evolution: One cannot be compare them with the data before evolution.
Hence, we used numerical evolution for some models.
 
To this end, we utilize simple cell models consisting of a large number of components and numerically evolve them under a given fitness condition to determine how the phenotypes of many components evolve. 
Two models are adopted in which phenotypes are generated by dynamical process for intracellular components. One is a catalytic reaction network model \cite{Zipf,SOC-Zipf} in which catalysts are synthesized with the aid of other catalysts so that the concentrations of a set of catalytic molecules constitute the phenotypic state space. The other model adopts a gene regulation network \cite{GRN1,GRN2,Wagner,KK-PLoS} in which proteins are expressed as a result of mutual activation or inhibition from other proteins. Both dynamics involve a large number of components, i.e., chemical concentrations or protein expression levels, which determine the phenotypes. The growth rate or fitness is determined by these phenotypes. 
 
In these models, genes govern the network structure and parameters for the reactions that establish the rules for such dynamical systems.
The phenotype of each organism, as well as the growth rate (fitness) of a cell, is determined by such reaction dynamics, whereas the evolutionary process consists of selection according to the associated fitness and genetic change in the reaction network (i.e., rewiring of the pathway). 
The global proportionality in concentration changes across components is confirmed in both the models after evolution.
 
Here, we explain the results of analysis using the catalytic network.
Despite its simplicity, this model captures the basic characteristic of cells such as the power-law abundance and log-normal fluctuations of cellular components, 
adaptation with fold-change detection, among other factors \cite{Zipf, Log-normal, SOC-Zipf, CFKK-JTB}.
 
In the model, the cellular state is represented by the numbers of $k$ chemical species, i.e., $(N_1,N_2,\cdots,N_k)$, whereas their concentrations are given by $x_i = N_i /V$ with the volume of the cell $V$ \cite{Zipf,SOC-Zipf}.
There are $m (<k)$ resource chemicals $S_1,S_2,\cdots,S_m$ whose concentrations in the environment and within a cell are given by $s_1,\cdots,s_m$ and $x_1,\cdots,x_m$, respectively. 
Each reaction leading from one chemical $i$ to another chemical $j$ was assumed to be catalyzed by a third chemical $\ell$, i.e., $i + \ell \rightarrow j + \ell$.
The resource chemicals are transported into the cell with the aid of other chemical components named as ``transporters.'' 
We assumed that the uptake flux of nutrient $i$ from the environment is proportional to $D s_i x_{t_i}$, where chemical $t_i$ acts as the transporter of nutrient $i$, and $D$ is a transport constant. For each nutrient, there is one corresponding transporter, represented by $t_i = m+i$.
The other $k-2m$ chemical species are catalysts synthesized from other components via the catalytic reactions mentioned above.
The catalytic reactions result in nutrient transformation into cell-component chemicals. With the uptake of nutrient chemicals from the environment, the total number of chemicals $N= \sum_i N_i$ in a cell increases. A cell, then, is divided into two cells when the total number of molecules exceeds a given threshold.
 
Here, to achieve a higher growth rate, the synthesis of the cell components must progress concurrently with nutrient uptake.
Hence, the cellular growth rate depends on the catalytic network, which is determined by genes. With evolution this growth rate, i.e., the fitness can be increased.
 
Because of this fitness, the evolutionary procedure is carried out as follows.
First, we prepared $n$ parent cells with slightly different reaction networks, randomly generated with a given connection rate. We applied stochastic reaction simulation of the above model and selected $n/L$ cells with high growth rates. From each of the $n/L$ parent cells, $L$ mutant cells were generated by replacing a certain fraction of reaction paths, whose rate is determined by the mutation rate. Next, we obtained $n$ cells of the next generation, which contained slightly different reaction paths. 
We repeated the same procedure to obtain the next generation population, and so on.
 
The simulation of evolutionary dynamics was performed under a constant environmental condition ${\bf o}=\{s^o_1,\cdots,s^o_m \}$. Under the original condition, the concentrations were set at $s^o_1=s^o_2=\cdots =s^o_m=1/m$, at which evolution progresses so that the cell growth rate, i.e., inverse of the average division time, is increased.

\subsection{Emergent global proportionality through evolution}

Using the model described in the last subsection, we analyzed the response of the component concentrations to the environmental change from the original condition. Here, the environmental condition is given by the external concentration $\{s_1,\cdots,s_m \}$. We then changed the condition to $s^{(\varepsilon,{\bf E})}_j=(1-\varepsilon)s^o_j+\varepsilon s^{\bf E}_j$, where $\varepsilon$ is the intensity of the stress and ${\bf E}=\{ s^{\bf E}_1,\cdots,s^{\bf E}_m \}$ denotes the vector of the new, stressed environment, in which the values of component $s^{\bf E}_1,\cdots,s^{\bf E}_m$ were determined randomly to satisfy $\sum_j s^{\bf E}_j =1$.
For each environment, we computed the reaction dynamics of the cell to obtain the concentration $x^{(\varepsilon,{\bf E} )}_j$ in the steady-growth state, from which the logarithmic change in the concentration $\delta X^{(\varepsilon,{\bf E} )}_j=\log (x^{(\varepsilon,{\bf E} )}_j/ x^{{\bf o}}_j)$ was obtained; the change in growth rate $\mu$, designated as $\delta \mu^{(\varepsilon,{\bf E} )}$, was also computed.
 
We next examined whether changes in $\delta X^{(\varepsilon,{\bf E} )}_j$ satisfy the common proportionality across all components under a variety of environmental changes for the ranges $0\leq \varepsilon \leq 1$. 
We examined the degree of proportionality both for the random networks before evolution and those after evolution under the given environmental conditions. We also tested whether the proportion coefficient is consistent with $\delta \mu^{(\varepsilon,{\bf E} )}$.
 
First, we computed the response of expression to the same type of stress, i.e., the same vector ${\bf E}$ with different intensities $\varepsilon$. As shown in Fig. 4(a), we examined the correlation between changes in component concentrations $\delta X^{(\varepsilon_1,{\bf E})}_j$ and $\delta X^{(\varepsilon_2,{\bf E})}_j$ caused by different magnitudes of environmental change ($\varepsilon_2 = \varepsilon_1 + \varepsilon$, at $\varepsilon >0$). For a small environmental change ($\varepsilon=0.02$), the correlation was sufficiently strong both for the random and evolved networks, whereas for a larger environmental change ($\varepsilon=0.08$), the correlation coefficients were significantly smaller for the random networks. We then computed the relationship between the ratio of the growth rate changes $\delta \mu^{(\varepsilon_1,{\bf E})}/\delta \mu^{(\varepsilon_2,{\bf E})}$ and fitted slope in $(\delta X^{(\varepsilon_1,{\bf E})}_j, \delta X^{(\varepsilon_2,{\bf E})}_j)$ across all components. Fig. 4(b) shows the ratio of the slope to $\delta \mu^{(\varepsilon_1,{\bf E})}/\delta \mu^{(\varepsilon_2,{\bf E})}$ (which turns to be unity when Eq. (3) is satisfied) as a function of the magnitude of environmental change $\varepsilon_1$. These results demonstrated that for the evolved network, Eq. (3) is maintained under large environmental changes, whereas it holds only against small changes for random networks. These results confirmed the expansion of the linear regime by evolution.

\begin{figure}[ht]
\begin{center}
\includegraphics[width=10cm]{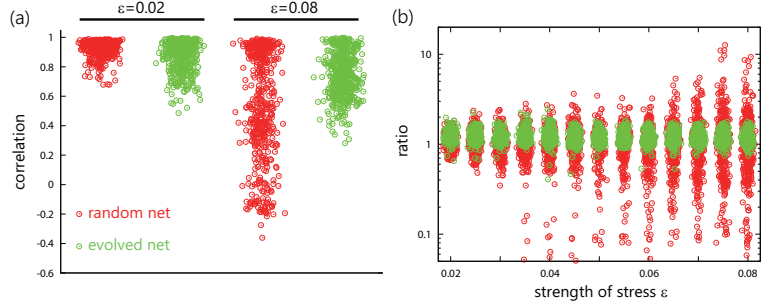}
\end{center}
\caption{
Common proportionality in concentration changes in response to the same type of stress in the catalytic-reaction network model.
(a) Coefficient of correlation between the changes in component concentrations $\delta X^{(\varepsilon_1,{\bf E})}_j$ and $\delta X^{(\varepsilon_2,{\bf E})}_j$ with $\varepsilon_2 = \varepsilon_1 + \varepsilon$. For the random and evolved networks, the correlation coefficients with a small ($\varepsilon=0.02$) and large ($\varepsilon=0.08$) environmental change are plotted, which were obtained using 100 randomly chosen environmental vectors ${\bf E}$. (b) Ratio of the slope in the concentration changes to the growth rate change as a function of the intensity of stress $\varepsilon$. The ratio in the ordinate becomes unity when Eq. (3) is satisfied. Reproduced from \cite{CFKK-Dominant}.
}
\end{figure}

Next, we examined the correlation of concentration changes under different types of environmental stressors. Fig. 5(a) shows examples of $(\delta X^{(\varepsilon,\bf{E}^a)}_j, \delta X^{(\varepsilon,\bf{E}^b)}_j)$ obtained by three networks from different generations. For the initial random networks, there was no correlation, whereas a modest correlation emerged in the 10th generation. Later, over evolution, common proportionality was observed; for instance, in the 150th generation, the proportionality reached more than two digits. To demonstrate the generality of the proportionality over a variety of environmental variations, we computed the coefficients of correlation between $\delta X^{(\varepsilon,\bf{E}^a)}_j$ and $\delta X^{(\varepsilon,\bf{E}^b)}_j$ for a random choice of different vectors $\bf{E}^a$ and $\bf{E}^b$. Fig. 5(b) shows the distributions of the correlation coefficients obtained by the random network and evolved network (150th generation).
Remarkably, global proportionality was observed even under different environmental conditions that had not been experienced through the course of evolution.
 
The relationship between $\delta \mu^{(\varepsilon_1,\bf{E}^a)}/\delta \mu^{(\varepsilon_2,\bf{E}^b)}$ and the slope in $(\delta X^{(\varepsilon_1,\bf{E}^a)}_j, \delta X^{(\varepsilon_2,\bf{E}^b)}_j)$ is presented in Fig. 5(c), whereas the ratio of the slope in $(\delta X^{(\varepsilon_1,\bf{E}^a)}_j, \delta X^{(\varepsilon_2,\bf{E}^b)}_j)$ to $\delta \mu^{(\varepsilon_1,\bf{E}^a)}/\delta \mu^{(\varepsilon_2,\bf{E}^b)}$ is shown in Fig. 5(d) as a function of $\varepsilon_2$. The slope of $\delta X$ agrees rather well with the growth rate change as given by Eq. (3) for the evolved networks as compared to the random networks.

\begin{figure}[ht]
\begin{center}
\includegraphics[width=10cm]{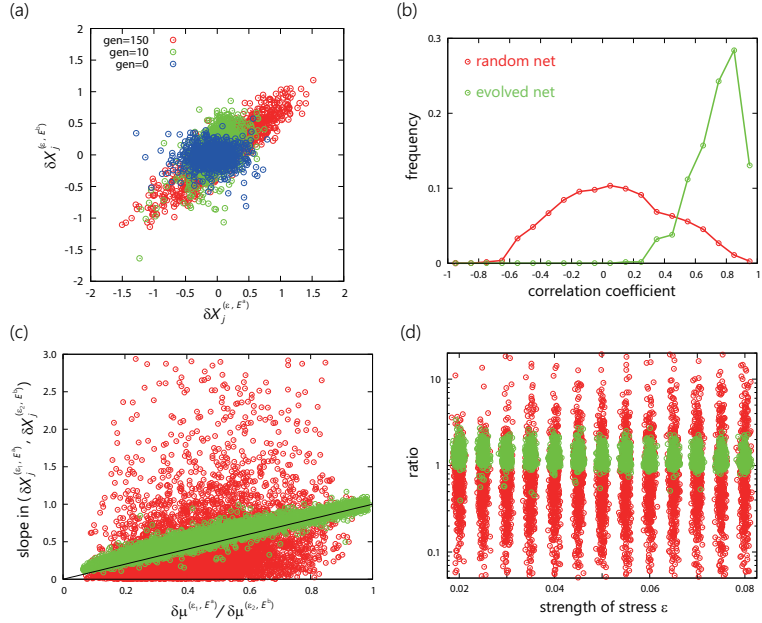}
\end{center}
\caption{
Common proportionality in concentration changes in response to different types of stress in the catalytic-reaction network model.
(a) Concentration changes across different types of environmental stressors.
For three different networks from different generations, $(\delta X_j({\bf E}^a), \delta X_j({\bf E}^b))$ are plotted by means of different randomly chosen vectors $\bf{E}^a$ and $\bf{E}^b$.
(b) Distributions of coefficients of correlation between the changes in component concentrations $\delta X^{(\varepsilon,{\bf E}^a)}_j$ and $\delta X^{(\varepsilon,{\bf E}^b)}_j$. Red and green curves represent the distributions of a random network and evolved network (150th generation) obtained from 1000 pairs of ${\bf E}^a$ and ${\bf E}^b$, respectively. The magnitude of environmental change $\varepsilon$ is fixed at $0.8$.
(c) Relation between the ratio of growth rate change $\delta \mu^{(\varepsilon_1,{\bf E}^a)}/\delta \mu^{(\varepsilon_2,{\bf E}^b)}$ and the slope in $(\delta X^{(\varepsilon_1,{\bf E}^a)}_j, \delta X^{(\varepsilon_2,{\bf E}^b)}_j)$.
(d) Ratio of the slope in the relation between concentration changes and growth rate change as a function of the intensity of stress $\varepsilon$ in the case of different types of stress. Reproduced from \cite{CFKK-Dominant}.
}
\end{figure}

The global proportionality over all components across various environmental conditions suggests that changes $\delta X^{(\varepsilon,{\bf E} )}_j$ across different environmental conditions are constrained mainly along a one-dimensional manifold after evolution has progressed (even) under a single environmental condition. To verify the existence of such constraints, we carried out the principal component (PC) analysis of the data of $\delta X^{(\varepsilon,{\bf E} )}_j$ across different environmental changes ${\bf E}$ and $\varepsilon$. As shown in Fig. 6(a), we plotted the data in the space with the first three PC axes. In the evolved network, high-dimensional data from $X^{({\varepsilon,{\bf E}})}_j$ were located along a one-dimensional curve.
Note that the contribution of the first PC reached 74\% in the data. In contrast, the data from the random network were scattered, and no clear structure was visible, as shown in Fig. 6(b).
Furthermore, for the evolved network, the value of the first PC agrees rather well with the growth rate \cite{CFKK-Dominant}.

\begin{figure}[ht]
\begin{center}
\includegraphics[width=9cm]{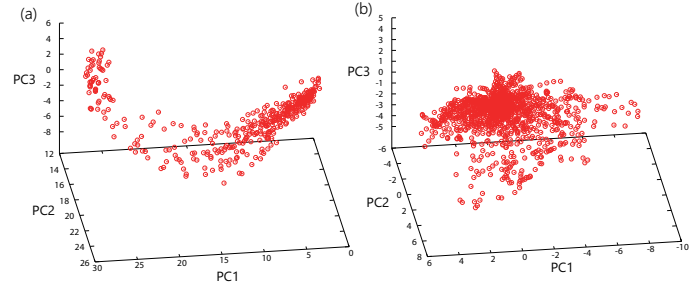}
\end{center}
\caption{
The change in $X^{(\varepsilon,{\bf E})}_j$ with environmental changes in principal-component space.
Component concentrations $X^{(\varepsilon,{\bf E})}_j$ at randomly chosen various ${\bf E}$ and $\varepsilon$ values are presented for (a) evolved and (b) random networks. In (a), the contributions of the first, second, and third components were 74\%, 8\%, and 5\%, respectively. Reproduced from \cite{CFKK-Dominant}.
}
\end{figure}

We then examined the evolutionary course of the phenotype projected on the same principal-component space, as depicted in Fig. 6(a). As shown in Fig. 7(a), the points from $\{ X_j \}$ generated by random mutations in the reaction network are again located along the same one-dimensional curve. Furthermore, those obtained by environmental variation or noise in the reaction dynamics also lie on this one-dimensional curve, as shown in Fig. 7(b). Thus, the phenotypic changes are highly restricted, both genetically and non-genetically, within an identical one-dimensional curve.
 As shown in Figs. 6 and 7, variation in the concentration due to perturbations was much larger along the first PC than along the other components. This suggests that relaxation is much slower in the direction of the first component than in the other directions. 
 
In summary, we observed emergent global proportionality which is far beyond the trivial linearity in response to tiny perturbations.
After evolution, the linearity region expanded to a level with an order-of-magnitude change in the growth rate. Additionally, the proportionality over different components across different environmental conditions was enhanced through evolution. In this global proportionality, evolutionary dimension reduction in phenotypic dynamics underlies changes in the high-dimensional phenotype space across a variety of environmental conditions, genetic variations, and noise, which are confined to a common one-dimensional manifold. The first principal-component mode corresponding to the one-dimensional manifold is highly correlated with the growth rate.

Notably, this dimension reduction by evolution has also been observed in some other models. First, even when the fitness for selection does not affect the growth rate but rather some other quantity (such as the concentration of a component), the phenotypic change is mainly constrained along a one-dimensional manifold. Second, even when the environmental condition (e.g., concentrations of external nutrients) is not fixed but rather fluctuates over generations, restriction within the one-dimensional manifold is observed \cite{TUSKK2019}. Third, there are preliminary reports that the evolution of gene regulation networks and spin-glass models \cite{Sakata} also show dimension reduction of phenotypic changes, as observed in our results.

\begin{figure}[ht]
\begin{center}
\includegraphics[width=9cm]{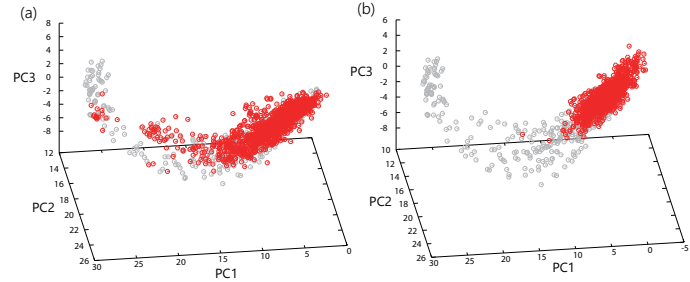}
\end{center}
\caption{
Change in component concentration $X_j$ because of (a) mutations and (b) noise in reaction dynamics. In (a), mutations were added to the evolved reaction network by randomly replacing 0.5\% of the reaction paths. The red dots show the concentrations of components after mutations, which are projected onto the same principal-component space, as depicted in Fig. 6(a). The gray dots represent the concentration changes caused by environmental changes for the reference, which are identical to those shown in Fig. 6(a). The red dots in (b) represent the concentration changes observed at each cell division, which are caused by the stochastic nature of reaction dynamics. Reproduced from \cite{CFKK-Dominant}.
}
\end{figure}

\section{Evolutionary Dimension Reduction Hypothesis}

Following the observation of global proportionality and dimension reduction from the high-dimensional phenotypic space in the last section, we proposed the following hypothesis: 

{\sl  
Phenotypic dynamics involve a large number of variables, and their state space is generally high-dimensional. However, phenotypic changes induced by environmental perturbations are constrained mainly along a low-dimensional (often one-dimensional) manifold (major axis). Along this manifold, the phenotypic dynamics are much slower than those across the manifold. Further, phenotypic changes induced by evolutionary changes (i.e., due to genetic changes governing phenotypic dynamics) also progress mainly along this manifold. The fitness gradually changes along the manifold, whereas it rapidly decreases across the manifold. 
} 

Indeed, the results described in the last section support the hypothesis, where the dominance of the first principal mode in a phenotypic change emerges after evolution, and the phenotypic changes as a result of mutation are constrained along the principal-mode axis (see Fig. 7). Moreover, expression data from bacterial evolution studies support this hypothesis, as described later.
 
This hypothesis is plausible considering the evolutionary robustness of phenotypes:
First, in most cases, phenotypes (e.g., concentrations of chemicals) are shaped as a result of complex dynamics involving a large degree of freedom. For examples, these dynamics can be determined by catalytic-reaction or gene-regulatory networks, which are determined by genes. In general, networks with higher fitness are rare, and thus a mutation-selection process is needed to achieve higher fitness. Because of the complexity of the dynamics, stochastic perturbations may influence the final phenotype in general, unless the networks are evolved to reduce the influence of perturbations \cite{ESB}.
 
As evolution progresses
robustness of the state against perturbations will be acquired. Otherwise, because of inevitable noise during the dynamics, a rare fitter state is not sustained. Increased robustness to perturbations is expected to result from evolution (see also \cite{Wagner,KK-PLoS}). Accordingly, in the state space, the dynamics provide flow to the selected (fitter) phenotype against perturbations for most directions, as shown schematically in Fig. 8. Strong contraction to the attracted state is shaped by evolution. However, there is (at least) one exceptional direction that does not possess such a strong contraction. This is the direction along which evolution to increase fitness progresses. Along this direction, phenotype states can be changed rather easily by perturbations. Otherwise, it is difficult for evolution to progress.
Hence, as schematically shown in Fig. 8, only along the direction in which evolutionary changes proceed, the relaxation is slow, whereas for other orthogonal directions, the change is much faster.

\begin{figure}[ht]
\begin{center}
\includegraphics[width=5cm]{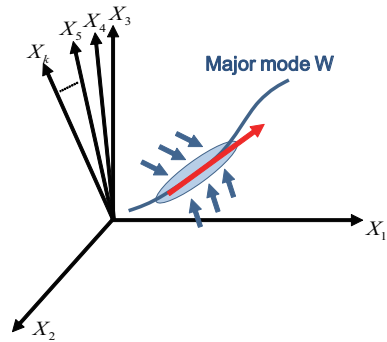}
\end{center}
\caption{
Schematic representation of the dimension-reduction hypothesis. In the state space of ${\bf X}$, dominant changes are constrained along the mode ${\bf W}$ following the major axis and its connected manifold, whereas the attraction to this manifold is much faster.
}
\end{figure}

Now, let us consider the relaxation dynamics to the original state (attractor) at a given generation, as represented by Eq. (2). The relaxation dynamics are represented by a combination of eigenmodes with negative eigenvalues. The magnitude of (negative) eigenvalues will be large, except for one (or a few) eigenvectors, whereas that along the direction of evolution will be much closer to zero; that is, relaxation along the direction of evolution is slower. Hence, variance along the largest principal component will be dominant, as demonstrated numerically in Fig. 6.
 
Indeed, in a recent study \cite{TUSKK2019}, the eigenvalues of the Jacobi matrix in Eq. (2) were numerically computed by using the catalytic-reaction dynamics adopted in the last subsection. The results confirmed that one eigenvalue was close to zero and all other (negative) eigenvalues had much larger magnitudes. This separation of one eigenvalue occurred because of evolution.
 
This hypothesis indicates that only one mode dominates in Eq.(2). Although the original dynamics are high-dimensional, most changes occur along the one-dimensional manifold ${\bf W}$, corresponding to the eigenvector for the smallest eigenvalue (or its nonlinear extension). Let us denote this direction as ${\bf w_0}$. 
From this dominance of the single dominant mode $W$, the global proportionality in Eq. (3) is naturally derived across different conditions ${\bf E}$. Because changes in all components $X_j$, $\delta X_j$s are constrained along the dominant mode $W$, they are given by the projection of the change in $W$ onto each $X_j$ axis, which is in turn given by $\cos \theta_j$, where $\theta_j$ is the angle between $W$ and each axis $X_j$. This angle is determined by the given phenotype state only and is independent of the type of environmental perturbation. Thus, the proportionality observed over different strengths of an identical environmental condition is valid across different types of environmental perturbations.

 Note that the change along ${\bf W}$ is parametrized by the growth rate $\delta \mu$ because it shows tight one-to-one correspondence with the principal coordinate. Then $\delta W$ is represented as a function of $\delta \mu$. Given that $\delta W \propto \delta \mu ({\bf E})$ for small $\delta \mu$, 
Then, Eq. (3) can be extended to different environmental vectors as follows:
\begin{equation}
\frac{\delta X_j({\bf E})}{\delta X_j({\bf E'})}= \frac{\delta\mu({\bf E})}{\delta\mu({\bf E'})}. 
\end{equation} 
Indeed, the above argument can be formulated explicitly by using linear algebra for the relaxation dynamics to the original state (attractor) at a given generation, as represented by Eq. (2). 
By using ${\bf L}={\bf J}^{-1}$, it follows that
\begin{math}
{\bf \delta X} = {\bf L}(\delta\mu{\bf I}-{\bf \gamma} \delta E),
\end{math} with ${\bf I}$ is an unit vector $(1,1,1,..1)^T$.
The matrix ${\bf L}$ is represented by 
$\sum_k \lambda_k {\bf w}_k{\bf v}_k^T$,
with their eigenvalues $\lambda_k$, and the corresponding right (left) eigenvectors ${\bf w}_k$ (${\bf v}_k$), respectively.
Noting that the hypothesis in the present section postulates that the magnitude of the smallest eigenvalue of ${\bf J}$ (denoted as $k=0$)
is much smaller than that of the others. In other words, the absolute eigenvalue of $\lambda_0$ for ${\bf L}={\bf J}^{-1}$ is much greater than others. Then, the major response to environmental changes, is given only by the projection to this mode for $\lambda_0$.
In other words, the major axis for the change ${\bf \delta X}$ is given by 
${\bf w_0}$. Using this reduction to this mode ${\bf w_0}$, and through straightforward calculation(see \cite{CFKK-Dominant} for details), we obtain ${\bf \delta X} = \lambda^0 {\bf w_0}(\delta \mu ({\bf v_0 \cdot I }))$.
Accordingly, the global proportionality relationship Eq. (5) is reproduced.
 
In summary, we can explain two basic features observed in experiments and simulations using the above theoretical formulation:
 
\noindent
 (1) {\bf Overall proportionality in expression level changes across most components and across various environmental conditions:} This is because high-dimensional changes are constrained to changes along the major axis, i.e., eigenvector ${\bf w_0}$.
 
\noindent
(2) {\bf Extended region of global proportionality:} Because the range in variation along ${\bf w_0}$ is large, the change in the phenotype is constrained to points near this eigenvector, causing the proportionality range of phenotypic change to extend via evolution. Furthermore, as long as the changes are nearly confined to the manifold along the major axis, global proportionality reaches the regime nonlinear to $\delta \epsilon$.

\section{Global Proportionality between Responses by Environmental and Evolutionary Adaptations}

According to the hypothesis in the last section, the change due to genetic variation would also be constrained along with this major axis ${\bf w}_0$, as the most changeable direction ${\bf w}_0$ is the direction in which evolution has progressed and will progress. Indeed, in the simulation described in Sec. 5, the phenotypic changes caused by the mutational change are constrained along the manifold spanned by the first principal mode in the environmentally induced phenotypic changes (see Fig.6). 
Accordingly, we expect to observe global proportionality between the concentration changes induced by a given environmental condition (stress) ($\delta X_j(Env)$) and those derived from evolution with genetic changes ($\delta X_j(Gen)$), as given by
\begin{equation}
\frac{\delta X_j (Gen)}{\delta X_j (Env)} =\frac{ \delta \mu (Gen)} {\delta \mu (Env)}. 
\end{equation}
 
For example, when cells are subjected to a stressed condition $Env$, the growth rate is reduced so that $\delta \mu (Env)<0$, whereas the expression levels change $\delta X_j(Env)$ accordingly. Next, the cells evolve under this given stressed condition over several generations along with genetic changes. After $n$ generations under genetic evolution, the growth rate recovers to some degree so that the growth rate shows a difference of $\delta \mu (Gen)$ from the original (no-stressed) state, satisfying $0 \geq \delta \mu(Gen) \geq \delta \mu(Env)$. The accompanied expression change, denoted by $\delta X_j(Gen)$, is then expected to satisfy
\begin{equation}
\frac{\delta X_j (Gen)}{\delta X_j (Env)} =\frac{ \delta \mu (Gen)} {\delta \mu (Env)} \leq 1
\end{equation} across most components $j$ in a similar manner as in Eq. (5).
Because $|\delta \mu (Gen)|$ is reduced with the progression of evolution, changes in the components introduced by the environmental change are reduced. Thus, there is an evolutionary tendency that the original expression pattern is recovered. This is reminiscent of the Le Chatelier principle in thermodynamics.
 
We next examined if the above relationship would hold in numerical simulation and bacterial evolution experiments.
 
\subsection{Verification by the reaction-network model}
 
We employed the catalytic reaction network model in Sec. 5.
After evolving the cell as described in the section under the given environmental condition, we switched the nutrient condition at a given generation. This caused the growth rate to decrease, which was later recovered through genetic evolution over generations. We computed the phenotypic changes induced by the environmental and evolutionary changes to examine the validity of the above relationship.

After altering the nutrient conditions, the abundances of all the components were changed. The average change of these abundances was denoted by:
$\delta X^{Env}_j \equiv \langle X_j(1)\rangle-\langle X_j(0) \rangle=\log \frac{\langle N_j(1) \rangle}{\langle N_j(0) \rangle}$, where generation 1 refers to the time point immediately after the environmental change, and generation 0 denotes the generation right before this nutrient change.
Similarly, we defined the response by genetic evolution after $m$ generations by $\delta X^{Gen}_j(m) = \langle X_j(m) \rangle- \langle X_j(0) \rangle$. Figs. 9(a) and (b) show the plot of
$\delta X^{Env}_j$ versus $\delta X^{Gen}_j(m)$ for $m=5$ and 50. The proportionality was observed between the environmental and genetic responses over all components.
 
Let us now define this proportion coefficient $r(m)$ for $\frac{\delta X^{Gen}_j(m)}{\delta X^{Env}_j}$ across components $j$. According to Eq. (6), this agrees with the growth rate change given by the ratio of $\delta \mu^{Gen}(m) =\mu(m)-\mu(0) (\leq 0)$ to $\delta \mu^{Env}=\mu(1)-\mu(0)(<0)$ at each generation $m$.
As shown in Fig. 9(c), the proportion coefficient $r(m)$ was plotted against this growth rate recovery $\delta \mu^{Gen}(m)/\delta \mu^{Env}$. 
The agreement between the two is clearly discernible.
 This proportion coefficient $r(m)$ is initially close to 1 (i.e., $m \sim 1$); with increasing generation $m$, the value decreased towards zero, in conjunction with recovery of the growth rate $\delta \mu^{Gen}(m)/\delta \mu^{Env}$. Thus, as stated in Le Chatelier principle mentioned in the previous section, evolution shows a common tendency to reduce changes in components introduced by environmental change.

\begin{figure}[ht]
\begin{center}
\includegraphics[width=10cm]{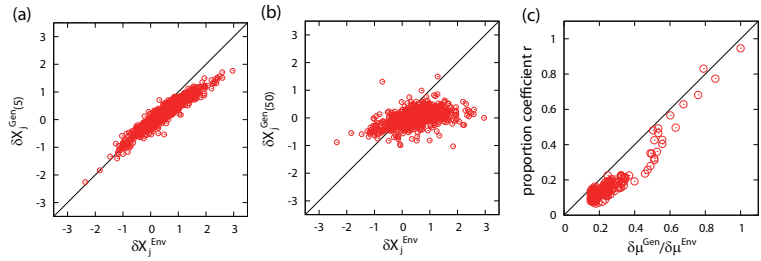}
\end{center}
\caption{
Response to environmental change versus response by evolution.
Relationship between the environmental response $\delta X^{Env}_j$ and genetic response $\delta X^{Gen}_j(m)$. (a) and (b) show the plots for $m=5$, and 50, respectively.
The black solid lines are $y=x$ for reference.
(c) Relationship between growth recovery rate $\delta \mu^{Gen}(m)/\delta \mu^{Env}$ and the proportion coefficient $r(m)$. The proportion coefficient $r(m)$ was obtained by using the least-squares method for the relationship of $\delta X^{Env}_j$ and $\delta X^{Gen}_j(m)$ for $m=1 \sim 200$. The black solid line is $y=x$ for reference. Reproduced from \cite{CFKK-Interface}.
}
\end{figure}
\subsection{Experimental confirmation by laboratory evolution}

To verify the relationship given by Eq.(7), we analyzed time-series transcriptome data obtained in an experimental evolution study of {\sl E. coli} under conditions of ethanol stress \cite{Horinouchi1, Horinouchi2}. In this experiment, after cultivation of approximately 1,000 generations (2,500 h) under 5\% ethanol stress, 6 independent ethanol-tolerant strains were obtained, which exhibited an approximately 2-fold increase in specific growth rates compared with the ancestor. For all independent culture series, mRNA samples were extracted from approximately $10^8$ cells at 6 different time points, and the absolute expression levels were quantified by microarray analysis. All mRNA samples were obtained from cells in an exponential growth phase, (see \cite{Horinouchi2} for details).
 
 Using the expression data taken at several generations through adaptive evolution, we analyzed the common proportionality in expression changes. The environmental response of the $j$-th gene $\delta X^{Env}_j$ is defined by the log-transformed ratio of the expression level of the $j$-th.
Similarly, the evolutionary response at $n$ hours after the exposure to stress $\delta X^{Gen}_j (n) $ is defined by the log-transformed ratio of the expression level at $n$ hours to that of the non-stress condition.
We found a common trend between the environmental and genetic responses over all genes \cite{CFKK-Interface}.
 
Furthermore, as shown in Fig. 10(a), the agreement between $r(n)$ and the growth recovery ratio $\delta \mu^{Gen}(n)/\delta \mu^{Env}$, as predicted by Eq.(7), was discernible, where $\delta\mu^{Gen}(n)$ and $\delta \mu^{Env}$ are the growth rate differences at $n$ h and 24 hours after the exposure to stress, respectively.
These results demonstrate that the evolutionary dynamics with growth recovery were accompanied by gene expression changes, which were reduced from those introduced by the new environment. 
 
How does $\{X_j \}$ change in the state space of a few thousand dimensions? As it is hard to see a high-dimensional state space, we used the first, second, and third principal components determined from the data for each generation of {\sl E. coli} gene expression. 
 (Approximately 31\% of the change in each data set can be explained by the 1st component, whereas 15\% is explained by the 2nd component). When the data points were plotted in the space of each principal component $P^i$-axis, they were distributed mainly along the $P^1$-axis direction, whereas the spread in the $P^2$ and $P^3$ axes was limited. Furthermore, the value of this first component was approximately proportional to the growth rate. This is consistent with the finding that the growth rate is a major factor in determining the change in expression of each gene.  Now, Fig.10(b) shows the cell-state changes by projecting the expression state $X_j$ at each generation onto these 3 component axes. Here, six independent data are superimposed, which were obtained by repeating the same experiment. Mutations occurred at different sites by each experiment; each of the six strains has a different genetic sequence. Nevertheless, all experimental samples changed with the same curve. The phenotypic changes to increase fitness are rather deterministic as compared to random changes in genetic sequences. Shaping the relevant phenotypic change is a priority in evolution, whereas several possibilities exist to achieve such changes genetically.

\begin{figure}[ht]
\begin{center}
\includegraphics[width=10cm]{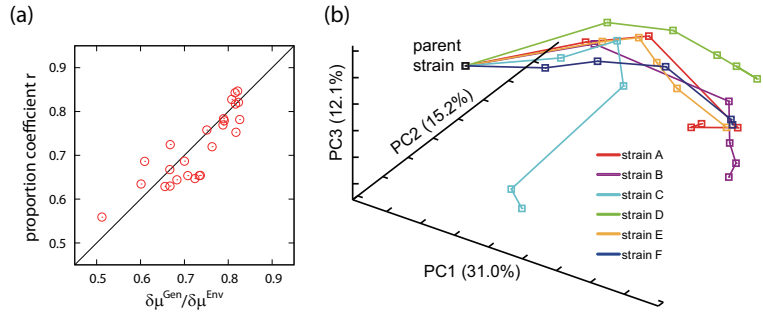}
\end{center}
\caption{
Response to environmental change versus response by evolution in {\sl E. coli} adaption to ethanol stress.
(a) Relationship between growth recovery rate $\delta \mu^{Gen}(n)/\delta \mu^{Env}$ and the proportion coefficient $r(n)$.
The proportion coefficient $r(n)$ was obtained by using the least-squares method for the relationship of $\delta X^{Env}_j$ and $\delta X^{Gen}_j(n)$ for $n=384$, 744, 1224, 1824, and 2496 hours.
The growth recovery rate $\delta \mu^{Gen}(n)/\delta \mu^{Env}$ was calculated based on the experimental measurements (see \cite{Horinouchi2} for details).
Among the 6 independent culture lines in \cite{Horinouchi2}, the results of 5 culture lines without genome duplication are plotted.
The black line is $y=x$ for reference.
(b) Changes in PCA scores during adaptive evolution. Starting from the parent strain, changes in the expression profiles during adaptive evolution are plotted as orbits in the three-dimensional PCA plane. Reproduced from \cite{CFKK-Interface}.
}
\end{figure}

\section{Evolutionary Fluctuation-Response Relationship}
 
In the previous section, we discussed the relationship between environmental and evolutionary responses. According to statistical physics, response and fluctuation are proportional. In evolution, analogously, we previously proposed an evolutionary fluctuation-response relationship: the phenotypic response throughout the evolutionary course is proportional to the phenotypic variance induced by noise \cite{Sato,ESB}. 
Consider a system characterized by a gene parameter $a$ and phenotypic trait $X$. We can then evaluate the change in $X$ against that in the gene parameter value from $a$ to $a+\Delta a$. Then, the proposed fluctuation-response relationship is given by
 
\begin{equation} \frac{\langle X \rangle_{a+\Delta a}- \langle X \rangle _{a}}{\Delta a} \propto \langle (\delta X)^2 \rangle,\end{equation}
 
\noindent
where $\langle X \rangle_a$ and $\langle (\delta X)^2 \rangle=\langle (X-\langle X \rangle )^2\rangle_a$ are the average and variance of the phenotypic trait $X$ for a given system parameterized by $a$, respectively. 
 
If $a$ is assigned as a parameter specifying the genotype (e.g., number of substitutions in the DNA sequence), this relationship implies that the evolutionary rate, i.e., change in average phenotype per generation, is proportional to the variance of the isogenic (clonal) phenotypic distribution, denoted by $V_{ip}=\langle (\delta X)^2 \rangle$. 
In fact, several model simulations and some experiments support this type of evolutionary fluctuation-response relationship \cite{Sato,CFKK-JTB}.
 
This evolutionary fluctuation-response relationship is associated with the phenotypic variance $V_{ip}$ of isogenic individuals, which is caused by noise. In standard population genetics, in contrast, the phenotypic variance due to genetic variation, named as genetic variance $V_g$ is considered. In fact, the so-called fundamental theorem of natural selection proposed by Fisher \cite{Fisher} states that the evolutionary rate is proportional to $V_g$. Thus, for both the evolutionary fluctuation-response relationship and Fisher's theorem to be valid, $V_{ip}$ and $V_g$ must remain proportional throughout the evolutionary course.
Indeed, such a relationship between these two variances was confirmed through evolutionary simulations of a catalytic reaction network model and gene regulation network \cite{CFKK-JTB,KK-PLoS}.
 
The origin of such a relationship can be explained as follows.
In general, developmental dynamics to shape the phenotype are quite complex, and the final state may be diverted by perturbations in the initial condition or by those that occur during the dynamics. 
Even if the fit phenotype is shaped by developmental dynamics, the perturbations due to noise during the dynamics may result in different non-fit states. Thus, the phenotype may be rather sensitive to noise.
After evolution progresses, the robustness of the fitted state to noise is increased. In this case, the global attraction to the target phenotype is shaped. This agrees with the hypothesis in Sec. 6.
 
Genetic changes, in contrast, can also cause perturbations to such dynamics. As the robustness to noise is shaped, the robustness against genetic changes is also expected to increase.
Through evolution, as the dynamics become increasingly robust to noise, they also become more robust to genetic changes, resulting in a correlation between the two types of robustness.
As the robustness to noise is increased, the phenotypic variance $V_{ip}$ will decrease; similarly, an increase in the robustness to genetic mutation leads to a decrease in $V_g$. Hence, throughout evolution, both $V_{ip}$ and $V_g$ decrease in correlation (or in proportion).
Thus, proportionality between $V_{ip}$ and $V_g$ is expected, as observed in the evolution simulations
\cite{CFKK-JTB,KK-PLoS}.
Furthermore, this proportionality of the two variances is extended to that between the phenotypic variance $X_i$
for each component $i$ by noise (written as $V_{ip(i)}$) and that by genetic variation (denoted by $V_g(i)$) \cite{CFKK-Interface,KK-JSP,ESB}. This, indeed, is expected by assuming that the relaxation of phenotypic changes is much slower along the dominant mode ${\bf W}$, phenotypic fluctuations due to noise are nearly confined to this axis. 

\section{Discussion}

In summary, we demonstrated the 2-by-2 global proportionality of phenotypic changes occurring between responses and fluctuations and between the perturbations due to environmental (noise) and genetic changes.
This proportionality is explained by evolutionary dimension reduction, which states that phenotypic changes due to environmental changes and genetic variation are constrained along a unique low-dimensional manifold, as observed in bacterial and numerical evolution experiments. 
Furthermore, expression changes induced by environmental stress, for most genes, are reduced through evolution to recover the growth, which is analogous to the Le Chatelier principle in thermodynamics \cite{Callen,Kubo}.
 
We demonstrated in numerical evolution that high-dimensional phenotypic change is mainly constrained along with the mode ${\bf w^0} $, the eigenvector corresponding to the eigenvalue of the relaxation dynamics closest to zero. The change in the phenotypic state is larger along the direction of ${ \bf w ^ 0} $, and the variable $W$ along this direction slowly returns to the steady state value. The time scale of this mode is distinctively slower than others, as confirmed directly by evolution simulation of the catalytic reaction and gene regulation networks \cite{TUSKK2019}.
This separation of the timescale of the slowest mode from others is theoretically expected in order to make the robustness of the fitted state and plasticity along the evolutionary course compatible.


Formation of one (or few) slow mode as given by $W$ separated from other modes is significant in evolutionary biology.
It may be possible that this type of mode is straightforwardly given by the expression of some specific genes that changes more slowly than others. However, it may be more natural that this $W$ is expressed as a collective change in the expressions of several genes rather than as a single protein. 
Because the slow mode is expressed by the first principal component, determination of genes whose expression levels contribute more to the first principal component will improve the understanding of how plasticity and robustness are compatible.
 
The slow, dominant mode $W$ emerges from evolution, but accelerates evolution. When faster and slower variables coexist and interact, the slower mode generally functions as a control variable for the faster variables. Accordingly, if the slower mode is modified by a genetic change, most faster variables will be influenced simultaneously. Furthermore, because the mode $W$ can influence the fitness $\mu$, the phenotypic evolution will be feasible simply by the change in this slow mode $W$. 
 
In contrast, if many variables change in a similar time scale, the genetic changes introduced to each will influence each other, and make directional phenotypic changes difficult to progress. This situation is reminiscent of the proverb ``Too many cooks spoil the broth.''
The emergence of slow modes governing the others has also been observed in the evolution of pattern formation \cite{Kohsokabe}.

The correlation between evolutionary and environmental responses raises a question regarding how the two processes with quite different time scales are correlated. The presence of the slow mode $W$ suggests a possible answer to this question. Adaptive dynamics, which originally show a much faster timescale than the evolutionary change, will be slowed along the mode $W$, whereas the evolutionary change, which originally has a much slower time scale is fastest, along the direction of the mode $W$.
Thus, along the dominant mode $W$, the timescales of phenotypic adaptation and evolution can approach each other.
 
Of course, further studies are needed to establish a phenomenological theory for phenotypic evolution. The generality of the evolutionary dimension reduction and resultant constraint in phenotypic evolution must be explored. The condition required for the emergence of dimensional reduction should also be determined. The models we studied satisfy the following conditions: (i) phenotypes with higher fitness are shaped by complex high-dimensional dynamical systems and (ii) the fraction of such fit states is rare in the state space and in the genetic-rule space. These two features are also consistent with our theoretical argument. The evolution of a statistical-physics model on interacting spins that preliminarily supports the dimension reduction also satisfies these conditions \cite{Sakata,Sakata2}. Studies of statistical physic and high-dimensional dynamical systems are needed to reveal the condition for evolutionary dimension reduction. Further experimental confirmation of the dimension reduction, as well as the directionality and constraint in phenotypic evolution, is needed, including in multicellular organisms.
 
Note that dimension reduction, or separation of a slow eigenmode from other faster modes, has been discussed in several other topics. They include protein dynamics \cite{Togashi,Tlusty}, laboratory ecological evolution \cite{Leibler}, learning in brain \cite{learning}, and neural networks \cite{Brenner}, among others. As a possible relationship, the {\sl sloppy parameter hypothesis} by Sethna \cite{sloppy} is proposed, which suggests that many parameters employed in biological models are irrelevant. Further studies are necessary to explain the universality of such evolutionary dimension reduction.

There are some limitations to dimension reduction. 
In the studies described here, we assumed steady-growth state, i.e., exponential phase. 
Under nutrient-limited conditions, however, there occurs 
a transition from such exponential growth to the stationary phase with suppressed growth, as has been  investigated both experimentally \cite{Balaban,stationary2} and theoretically \cite{Himeoka,Dill}. As such phase with suppressed growth may not be selected as a robust fitted state, whether the dimension reduction is valid to it requires further analysis.
~\\

{\bf Acknowledgment}

The authors would like to thank Takuya U Sato and Tetsuhiro Hatakeyama for stimulating discussions.
This research was partially supported by a Grant-in-Aid for Scientific Research (S) (15H05746) from the Japanese Society for the Promotion of Science (JSPS) and Grant-in-Aid for Scientific Research on Innovative Areas (17H06386 and 17H06389) from the Ministry of Education, Culture, Sports, Science and Technology (MEXT) of Japan.

\end{document}